\documentstyle[12pt]{article}
\begin{document}
\tolerance=5000
\def\pp{{\, \mid \hskip -1.5mm =}}
\def\cL{{\cal L}}
\def\be{\begin{equation}}
\def\ee{\end{equation}}
\def\bea{\begin{eqnarray}}
\def\eea{\end{eqnarray}}
\def\tr{{\rm tr}\, }
\def\nn{\nonumber \\}
\def\e{{\rm e}}
\def\D{{D \hskip -3mm /\,}}

\def\SEH{S_{\rm EH}}
\def\SGH{S_{\rm GH}}
\def\AdS5{{{\rm AdS}_5}}
\def\S4{{{\rm S}_4}}
\def\gfv{{g_{(5)}}}
\def\gfr{{g_{(4)}}}
\def\SC{{S_{\rm C}}}
\def\RH{{R_{\rm H}}}

\def\wlBox{\mbox{
\raisebox{0.1cm}{$\widetilde{\mbox{\raisebox{-0.1cm}\fbox{\ }}}$}}}
\def\htBox{\mbox{
\raisebox{0.1cm}{$\hat{\mbox{\raisebox{-0.1cm}{$\Box$}}}$}}}

\  \hfill 
\begin{minipage}{3.5cm}
May 2001 \\
\end{minipage}

\vfill

\begin{center}
{\large\bf Cosmological constant and Brane New World}

\vfill

{\sc Shin'ichi NOJIRI},\footnote{email: nojiri@cc.nda.ac.jp} 
{\sc Octavio OBREGON}$^{\spadesuit}$\footnote{email:octavio@ifug3.ugto.mx}
and {\sc Sergei D. ODINTSOV}$^{\spadesuit}$\footnote{
On leave from Tomsk State Pedagogical University, RUSSIA. \\
\ \hskip 1cm email: odintsov@ifug5.ugto.mx}, \\

\vfill

{\sl Department of Applied Physics \\
National Defence Academy, 
Hashirimizu Yokosuka 239, JAPAN}

\vfill

{\sl $\spadesuit$ 
Instituto de Fisica de la Universidad de 
Guanajuato \\
Apdo.Postal E-143, 37150 Leon, Gto., MEXICO}

\vfill


{\bf ABSTRACT}

\end{center}
 The estimation of the cosmological constant in inflationary 
 Brane New World models is done.
It is shown that basically it is quite large, of the same order as in 
anomaly-driven inflation. However, for some fine-tuning of bulk gravitational 
constant and AdS scale parameter $l^2$ it maybe reduced to sufficiently small 
value. Bulk  higher derivative AdS gravity with quantum brane matter
 may also serve as the  model where  
small positive cosmological constant occurs.



\newpage

\section{Introduction.}

It is a quite well-known fact that energy density of the vacuum appears 
in Einstein equations in the form of an effective cosmological constant.
In other words, vacuum (or vacuum polarization) induces 
the effective cosmological constant which curves the observable 4d world.
Roughly speaking, 4d curvature is of the order of the square root 
from the cosmological constant which should include 
not only vacuum contribution but also other (dark?) matter contributions.
According to recent observations (for a review and list of 
references, see
\cite{C,W}) the cosmological constant is positive and small.
It is interesting that positive small cosmological constant is not
 what is expected from string theory.  Nevertheless, there are some 
suggestions how to get small cosmological constant within string theory
 (see for example, refs.\cite{McI,CL} what maybe related with wormholes or spacetime foam \cite{R}). 

The fundamental question in cosmology is why the observable cosmological
 constant is so small? QFT considerations predict quite large vacuum energy
and hence, quite large cosmological constant.
Of course, it could be that the cosmological constant at very early Universe 
was large. However, due to some dynamical mechanism (supersymmetry?
orbifold compactification?) it was reduced to the current small value.
It would be interesting to understand the role of quantum effects 
as concerns to cosmological constant in brane-world physics.
In the present work we discuss the cosmological constant value
which  appears 
in Brane New World suggested in refs.\cite{HHR,NOplb}. Brane New World 
scenario represents quantum (or AdS/CFT induced) generalization 
of Randall-Sundrum Universe 
\cite{RS} where brane quantum fields are taken into account. It is interesting 
that quantum-induced brane inflation \cite{HHR, NOplb} 
(for related works, see \cite{other}) 
 occurs in the analogy 
with trace-anomaly driven inflation \cite{sta}.

\section{Quantum-corrected cosmological constant}

We start from the FRW-universe equation of motion 
with quantum corrections (taking into account 
conformal anomaly-induced effective action). 
 Such quantum-corrected FRW-equation has the form \cite{NO}:
\bea
\label{od1}
H^2&=& - {1 \over a^2} + {8\pi G \over 3}{E \over V} \nn
&& + {8\pi G \over 3}\left[
 - b'\left(4H H_{,tt} + 12 H_{,t} H^2 - 2H_{,t}^2 + 6H^4 
+ {8 \over a^2} H^2\right) \right.\nn
&& +{1 \over 12}\left\{b'' + {2 \over 3}\left(b+b'\right)
\right\} \nn
&& \quad \times \left(- 36 H_{,t}^2 + 216 H_{,t} H^2 
+ 72H H_{,tt} - {72 \over a^2}H^2+{36 \over a^4} \right) \nn
&& \left. + {\tilde a \over a^4}\right] \ ,
\eea
where $V$ is the spatial volume of the universe, 
$\tilde a=-8b'$ (a normalization choice \cite{NO}), $b''=0$,
$b$ is not necessary in the subsequent analysis and 
\bea
\label{Ivii}
b'&=&-{N+11N_{1/2}+62N_1 -28 N_{\rm HD} + 1411 N_2 
+ 1566 N_W \over 360(4\pi)^2}\ .
\eea
Here $N$, $N_{1/2}$, $N_1$, $N_{\rm HD}$ 
are the number of scalars, (Dirac) spinors, vectors and higher 
derivative conformal scalars which are present in conformal QFT
 filling the Universe. 
The quantity $N_2$ denotes the contribution to conformal 
anomaly from a spin-2 field 
(Einstein gravity) and $N_W$ the contribution from 
higher-derivative Weyl gravity. 
As usually, the quantum corrections produce 
an effective cosmological constant.
 
In the absence of classical matter energy ($E=0$), 
the general FRW equation 
allows the quantum-induced de Sitter space solution (anomaly-driven inflation 
\cite{sta}):
\be
\label{od2}
a(t)=A\cosh Bt\ ,\quad ds^2= dt^2 
+ A^2 \cosh^2{t \over A}d\Omega_3^2\ ,
\ee
where $A$ is a constant and
$B^2={1 \over A^2}= - {1 \over 16\pi Gb'}$. 
It is evident then that 
 the effective cosmological 
constant is defined as follows
\be
\label{cscn1}
\Lambda_{\rm eff}={3 \over A^2}
= - {3 \over 16\pi Gb'}\ .
\ee
If $b'$ is of order unity (what is typical in Standard Model) , we find 
\be
\label{cscn1b}
\Lambda_{\rm eff}\sim\left(10^{19}{\rm GeV}\right)^2\ ,
\ee
  It is quite large. 

The natural question now is: what happens for similar inflationary brane-world 
scenario? Following the approach of ref.\cite{SV} in 
 \cite{NObr},  
the quantum-corrected FRW-type equation was considered 
as it is predicted by inflationary 
 brane universe in the bulk 
Schwarzschild-AdS$_5$ spacetime:
\be
\label{AdSS} 
ds_{\rm AdS-S}^2 = {1 \over h(a)}da^2 - h(a)dt^2 
+ a^2 d\Omega_3^2 \ ,\ \ 
h(a)= {a^2 \over l^2} + 1 - {16\pi G_5M \over 3 V_3 a^2}\ .
\ee
Here $G_5$ is 5d Newton constant and 
$V_3$ is the volume of the unit 3 sphere.  
The quantum-corrected FRW type equation looks as \cite{NObr}
\bea
\label{e10}
&& H^2 = - {1 \over a^2} 
+ {8\pi G \rho \over 3} \\
\label{e10b}
&& \rho={l \over a}\left[ {M \over V_3 a^3} \right. 
+ {3a \over 16\pi G_5}\left[
\left[{1 \over l} + {\pi G_6 \over 3}\left\{ 
-4b'\left(\left(H_{,\tilde t \tilde t \tilde t} + 4 H_{,\tilde t}^2 
+ 7 H H_{,\tilde t\tilde t} \right.\right.\right.\right. \right.\nn
&& \ \left.\left. + 18 H^2 H_{,\tilde t} + 6 H^4\right) 
+ {4 \over a^2} \left(H_{,\tilde t} + H^2\right)\right) 
+ 4(b+b') \left(\left(H_{,\tilde t \tilde t \tilde t} 
+ 4 H_{,\tilde t}^2 \right. \right. \nn
&& \ \left.\left.\left.\left.\left.\left. + 7 H H_{,\tilde t\tilde t} 
+ 12 H^2 H_{,\tilde t} \right) - {2 \over a^2} 
\left(H_{,\tilde t} + H^2\right)\right) \right\}\right]^2
 - {1 \over l^2} \right]\right]\ .
\eea
Here 4d Newton constant $G$ is given by
\be
\label{e12}
G={2G_5 \over l}\ .
\ee
When $M=0$, the above equation (\ref{e10}) has a solution of the 
form (\ref{od2}) if $A^2={1 \over B^2}$ and 
\be
\label{dS3} 
0= - B^2 - {1 \over l^2} 
+ \left({1 \over l} - 8\pi G_5 b' B^4 \right)^2 \ .
\ee
 For negative $b'$ (\ref{dS3}) has a unique solution. 
The solution is nothing but the de Sitter brane solution 
in \cite{HHR,NOplb}. 
Eq.(\ref{dS3}) can be rewritten as
\be
\label{cscn01}
0=-1 + 2\beta C + \beta^2 C^3\ .
\ee
Here 
\be
\label{cscn02}
C\equiv l^2 B^2\ ,\quad 
\beta = - {8\pi G_5 b'\over l^3} = - {4\pi G b' \over l^2} \ .
\ee
Then the effective cosmological constant is given by
\be
\label{cscn2}
\Lambda_{\rm eff}={3 \over A^2}=3B^2 = {3C \over l^2}\ .
\ee
If $|\beta|\gg 1$,  a solution of (\ref{cscn01}) 
is given by
\be
\label{cscn3}
C={1 \over 2\beta}
\left( 1 + {\cal O}\left( {1 \over \beta} \right) \right)
= - {l^2 \over 8\pi G b'}
\left( 1 + {\cal O}\left( {1 \over \beta} \right) \right)\ ,
\ee
and one gets 
\be
\label{cscn4}
\Lambda_{\rm eff}= - {3 \over 8\pi G b'}
\left( 1 + {\cal O}\left( {1 \over \beta} \right) \right)\ ,
\ee
which is different from (\ref{cscn1}) by factor two but 
there is no qualitative difference. 
On the other hand, if $|\beta|\ll 1$, a solution of (\ref{cscn01}) 
is given by
\be
\label{cscn5}
C={1 \over \beta^{2 \over 3}}
\left( 1 + {\cal O}\left( \beta^{1 \over 3} \right) \right)
= \left(-{l^2 \over 4\pi G b'}\right)^{2 \over 3}
\left( 1 + {\cal O}\left( \beta^{1 \over 3} \right) \right)\ ,
\ee
and we find
\be
\label{cscn6}
\Lambda_{\rm eff}= - {3 \over l^2}
\left(-{l^2 \over 4\pi G b'}\right)^{2 \over 3}
\left( 1 + {\cal O}\left( \beta^{1 \over 3} \right) \right)\ .
\ee
Since $|\beta|\ll 1$ means $l^3 \gg G_5$ or $l^2 \gg G$, 
the effective cosmological constant $\Lambda_{\rm eff}$ can be 
small in this case. Note that one can write other solutions for
effective cosmological constant from above cubic equation. 
However, in most cases it is getting very large.

 Motivated  by AdS/CFT correspondence (for a review, see\cite{AdS}), we 
may consider ${\cal N}=4$ $SU(N)$ SYM theory on the brane. 
Then 
\be
\label{N4}
b=-b'={N^2 -1 \over 4(4\pi )^2}
\ee
and 
\be
\label{AdSCFT}
{l^3 \over G_5}={2N^2 \over \pi}\ .
\ee
In the large $N$ limit, we have 
\be
\label{betaN4}
\beta = - {8\pi G_5 b'\over l^3} ={1 \over 16}\ .
\ee
Then by numerical solving (\ref{cscn01}), one finds
\be
\label{CN4}
C=4.71804\ .
\ee
 In this case $\Lambda_{\rm eff}\sim {\cal O}\left(l^{-2}\right)
\sim \left(10^{19}{\rm GeV}\right)^2$ again. Hence, for decreasing 
 the cosmological constant one has to consider QFT which is not exactly 
conformally invariant (for a recent AdS/CFT discussion of such theories, see 
\cite{MPV}).
 From another point, one may take the arbitrary 
bulk values for AdS parameter and five-dimensional gravitational constant 
in order to achieve the smallness of the cosmological constant. 
The drawback of this
is evident: it is kind of fine-tuning. 

If we include quantum bulk scalar or spinor , they 
induce the Casimir effect in orbifold compactification. 
The corresponding vacuum energy which may stabilize the radius was found 
 in \cite{NOZcs,casimir}. 
In the Euclidean signature, de Sitter space is expressed as 
the sphere. In \cite{NOZcs}, the Casimir effect makes the 
radius smaller or larger, especially the conformal scalar in the 
bulk makes the radius small and time-dependent. In 
Minkowski signature, the inverse of the radius corresponds 
to the expansion rate (i.e. $B$) of the universe. 
 Then from Eq.(\ref{cscn2}), the conformal 
scalar in the bulk increases the effective cosmological constant.
Note, however, that taking into account the bulk quantum gravity 
with 5d cosmological constant may presumably help in resolution of the problem.
Unfortunately, the corresponding calculation is quite complicated 
and it is not done so far.

 One may consider 5d $R^2$ gravity, whose action is given by:
\be
\label{vi}
S=\int d^5 x \sqrt{-\hat G}\left\{a \hat R^2 
+ b \hat R_{\mu\nu}\hat R^{\mu\nu}
+ c \hat R_{\mu\nu\xi\sigma}\hat R^{\mu\nu\xi\sigma}
+ {1 \over \kappa^2} \hat R - \Lambda \right\}\ , 
\ee
and the brane with quantum  matter corrections 
as in the previous case. 
When $c=0$, the net effects can be absorbed into the 
redefinition of the Newton constant and the length 
scale $l$ of the AdS$_5$ \cite{NOOr2}, given by 
\bea
\label{r2a1}
&& {1 \over 16\pi G_5}={1 \over \kappa^2} 
\rightarrow {1 \over 16\pi \tilde G_5}
={1 \over \tilde\kappa^2} 
={1 \over \kappa^2} - {40a \over l^2}  - {8b \over l^2}\ ,\\
\label{ll4}
&& 0={80 a \over l^4} + {16 b \over l^4} 
 - {12 \over \kappa^2 l^2}-\Lambda\ .
\eea
Then  assuming the brane solution as in (\ref{od2}) ( Brane New World 
in higher derivative gravity),  one obtains the 
analogue of (\ref{dS3}) from corresponding FRW-equation
\be
\label{dS3b} 
0= - B^2 - {1 \over l^2} 
+ \left({1 \over l} - 8\pi \tilde G_5 b' B^4 \right)^2 \ .
\ee
Furthermore if we replace $G$ in (\ref{e12}) by 
\be
\label{e12b}
G={2\tilde G_5 \over l}\ ,
\ee
the arguments from (\ref{cscn3}) to (\ref{cscn6}) are valid. 
In other words, using hidden parameters of bulk higher derivative terms
one can obtain the 4d cosmological constant to be reasonably small.
This picture maybe generalized for the case of non-zero $c$, however,
the corresponding equation for cosmological constant is more complicated.
Nevertheless, the qualitative conclusions will be the same.

\section{Discussion}

In summary, we considered the effective cosmological constant in Brane 
New World induced by quantum brane matter effects. Rough estimation 
for inflationary brane indicates that in most cases the cosmological constant 
is quite large. Fine-tuning of bulk 5d gravitational and 5d cosmological
constant may sometimes lead to significant decrease of 4d cosmological constant.
Of course, one can imagine the situation that large cosmological constant 
at the beginning of inflationary era is reduced to current small value 
by some mechanism in course of evolution. Nevertheless, it looks that 
 New Brane World scenario by itself cannot suggest a natural way to solve
 the cosmological constant problem.

\end{document}